\documentclass[aps,twocolumn,superscriptaddress,floatfix,nolongbibliography,noeprint]{revtex4-2}
\usepackage{algorithm}
\usepackage{algpseudocode}
\usepackage{blindtext}
\usepackage{bm}
\usepackage{braket}
\usepackage[colorlinks,bookmarks=false,citecolor=teal,linkcolor=red,urlcolor=blue]{hyperref}
\usepackage{amssymb}
\usepackage{amsmath}
\usepackage{graphicx}
\usepackage{float}
\usepackage{bbold}
\usepackage{xcolor}
\usepackage{mathtools}
\usepackage{amsthm}
\algrenewcommand\algorithmicrequire{\textbf{Input:}}
\algrenewcommand\algorithmicensure{\textbf{Output:}}
\newcommand{\I}{\mathrm{i}}
\newcommand{\id}{\mathbb{1}}

\renewcommand{\t}[1]{\textrm{#1}}
\newcommand{\Tr}{\mathrm{Tr}}
\newcommand{\T}{\mathrm{T}}
\newtheorem{theorem}{Theorem}

\theoremstyle{definition}
\newtheorem{definition}[theorem]{Definition}

\begin{document}
\title{Adaptive quantum channel discrimination using methods of quantum metrology}
\author{Stanisław Sieniawski}
\email[Corresponding author: ]{s.sieniawski@student.uw.edu.pl}
\affiliation{Faculty of Physics, University of Warsaw, Pasteura 5, 02-093 Warszawa, Poland}
\author{Rafał Demkowicz-Dobrzański}
\email{demko@fuw.edu.pl}
\affiliation{Faculty of Physics, University of Warsaw, Pasteura 5, 02-093 Warszawa, Poland}

\begin{abstract}
    We present an efficient tensor-network based algorithm for finding the optimal adaptive quantum channel discrimination strategies inspired by recently developed numerical methods in quantum metrology to find the optimal adaptive channel estimation protocols.
    We examine the connection between 
    channel discrimination and estimation problems, highlighting in particular
    an appealing structural similarity between models that admit Heisenberg scaling estimation performance, and models that admit perfect channel discrimination in finite--number of channel uses. 
\end{abstract}

\maketitle

\section{Introduction}
\label{sec:introduction}

The question of statistical distinguishability of quantum objects arises naturally from experiment.
One basic, but already surprising, instance is the task of quantum state discrimination.
First work on this problem was done in \cite{helstromQuantumDetectionEstimation1969,holevoStatisticalDecisionTheory1973}.
Importantly, two non-orthogonal quantum states cannot be distinguished perfectly using any finite number of copies of the states \cite{audenaertDiscriminatingStatesQuantum2007}.
Closely connected to this task is the discrimination of quantum channels.

Quantum channels are descriptions of evolutions of quantum physical systems.
Discrimination of quantum channels is a mathematical model of experimental discrimination between different evolutions quantum systems undergo.
Fundamental study of this problem is interesting as it may yield bounds on experimental capacity, but may also suggest how to construct optimal experimental setups.
The problem has been studied thoroughly.
From discrimination of unitary channels \cite{acinStatisticalDistinguishabilityUnitary2001}, through the question of perfect distinguishability with finite number of channel uses \cite{duanPerfectDistinguishabilityQuantum2009} to the question of the hierarchy of parallel, adaptive and indefinite causal order strategies \cite{chiribellaMemoryEffectsQuantum2008,harrowAdaptiveNonadaptiveStrategies2010,krawiecDiscriminationPOVMsRankone2020,bavarescoStrictHierarchyParallel2021,bavarescoUnitaryChannelDiscrimination2022, WinterAdaptive2022}.

In channel discrimination problems, one typically chooses the average success probability over some ensemble of channels as the figure-of-merit, see \cite{rexitiDiscriminationDephasingChannels2022,soedaOptimalQuantumDiscrimination2021,maciejewskiOperationalQuantumAverageCase2023, gutoskiMeasureDistanceQuantum2012}.
Usefulness of different quantum resources for discrimination was also studied \cite{urrehmanQuantumChannelDiscrimination2018,oskoueiProfitableEntanglementChannel2023} and bounds for different models were developed \cite{Pirandola2019, pereiraBoundsAmplitudeDamping2021,itoLowerBoundsError2022,nakahiraSimpleUpperLower2021,zhuangUltimateLimitsMultiple2020}.
Interesting related problems and generalisations include the tasks of measurement discrimination \cite{dattaPerfectDiscriminationQuantum2021, Krawiec2024discrimination} and quantum network discrimination \cite{hircheQuantumNetworkDiscrimination2023,nakahiraQuantumProcessDiscrimination2021}.

The problem of finding the optimal adaptive strategy for discrimination can be solved explicitly as a semidefinite program (SDP) \cite{Katariya_2021, gutoskiMeasureDistanceQuantum2012}.
The problem, however, scales exponentially with number of channel-uses. Hence, in order to analyse the problem in the large number of uses regime, we need another optimisation method. 

In this paper we present a tensor-network based algorithm that allows to optimize effectively adaptive channel discrimination strategies, involving multiple channel-uses,
written in the language of quantum combs \cite{chiribellaTheoreticalFrameworkQuantum2009}.
The algorithm presented here is an adaptation of the tensor-network algorithm for quantum channel estimation designed in  \cite{kurdzialekQuantumMetrologyUsing2025} and implemented in the 
QMetro++ package \cite{dulianQMetroPythonOptimization2025}. 
The main difference between the formulation of the two algorithms  is the form of input and the optimized figure-of-merit. Since the channel estimation problem is formulated within the local frequentist estimation paradigm \cite{Demkowicz2015}, we are given a channel as well as its parameter derivative at some fixed parameter value, and optimize the corresponding quantum Fisher information (QFI).  In a channel discrimination problem, we are given a set of channels together with prior probabilities (a Bayesian framework) and maximize the average probability of success.

The developed optimisation algorithm returns a mathematical description of the supposed optimal strategy---thus formally giving a lower bound for probability of success for the truly optimal strategy. In order to benchmark its optimality, we also compare the results obtained with the known upper bounds. We will use bounds that are derived directly for a given channel discrimination problem, as in e.g. unitary channel discrimination case \cite{acinStatisticalDistinguishabilityUnitary2001}, or if the relevant bounds are not known, use a connection between the discrimination problem and the corresponding estimation problem \cite{albarelliProbeIncompatibilityMultiparameter2022}, to bound the discrimination error in terms of QFI bounds for a corresponding channel estimation problem, which we numerically compute using again the QMetro++ package \cite{dulianQMetroPythonOptimization2025}.

\begin{figure*}[t]
    \centering
    \includegraphics[scale=0.8]{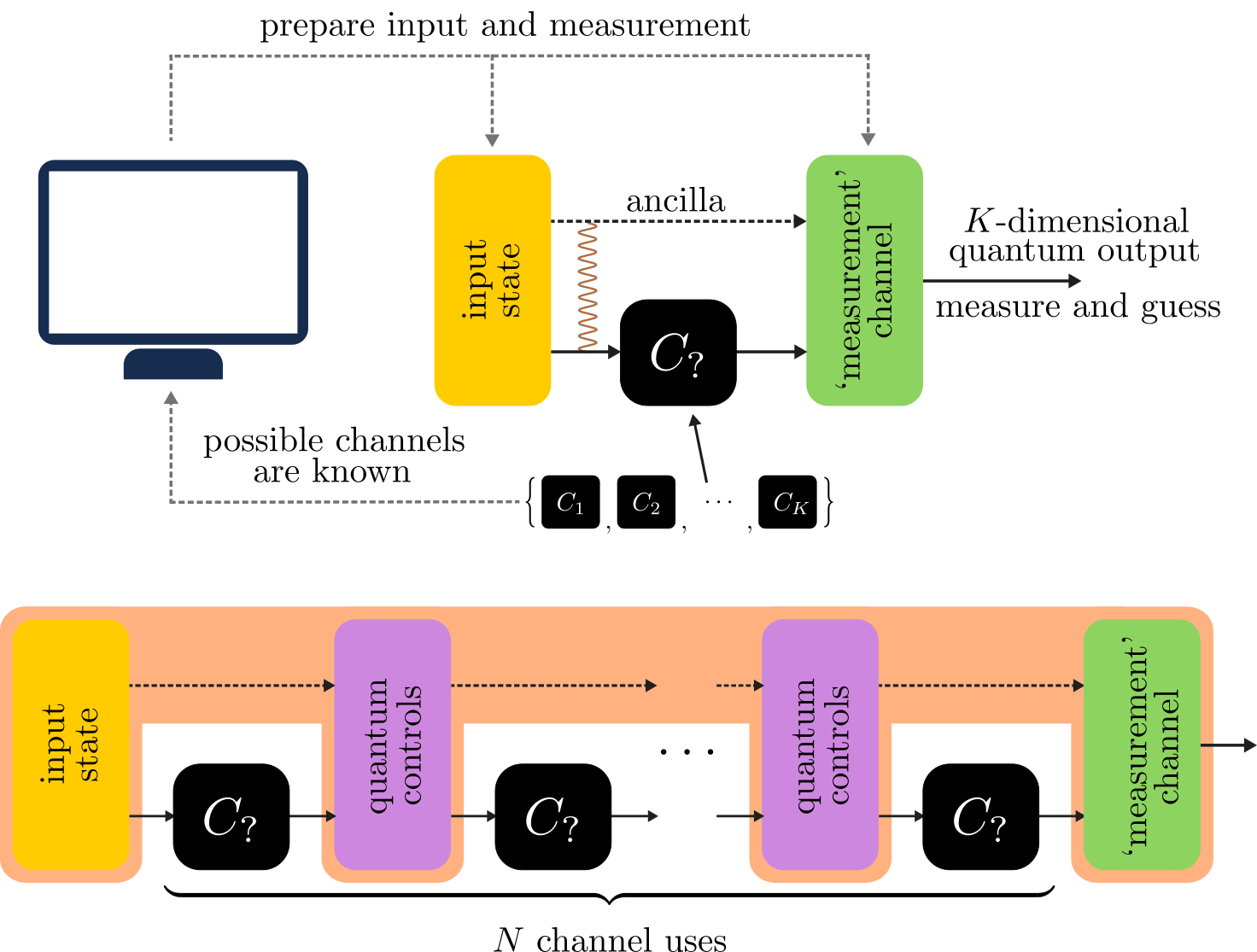} 
    \caption{Simple diagram representing the task of quantum channel discrimination with single use of the channel.
    A player (depicted here as the computer) knows the forms of the possible quantum channels from which she has been given one randomly, according to a distribution she also knows.
    Her task is to choose the state that she inputs into the channel (possibly entangled with some ancillary space) and measurement scheme at the output of the channel that maximises her probability of successfully guessing which channel she has.
    The lower part represents an adaptive strategy for multiple channels uses, where additional quantum controls are allowed between subsequent channel-uses.
    The light orange shade represents the quantum comb --- mathematical representation of the strategy.}
    \label{fig:intro}
\end{figure*}

\section{Optimal quantum channel discrimination}

Let us describe the task of quantum channel discrimination as the following game.
A player is given a channel $C_k$ with probability $p_k$ from an ensemble of $K$ channels (she knows the values of $\{p_k\}$ and the forms of $\{C_k\}$) and she is allowed $N$ uses of the channel.
Her task is to construct the optimal discrimination strategy: prepare a quantum state, send it through the channel $N$ times, possibly transforming the state between uses, perform a measurement and depending on the result guess which $C_k$ she got, maximising her probability of success (or, equivalently, minimising probability of error).
We assume that input and output spaces of all the channels have the same dimension.
The task is illustrated in Fig.~\ref{fig:intro}.

Here we allow adaptive strategies modelled mathematically as quantum combs.
Adaptive means that the player is allowed to transform the state between following uses of the channel
(footnote \footnote{There is, in principle, a more general construction of quantum experiment strategies.
One could, for example, interact with two uses of a channel in a temporal order that is controlled by some ancillary quantum state (e.g. $|0\rangle$ interact with one use first, $|1\rangle$ interact with the other first).
For some channel discrimination problems, it has been shown that this generalisation yields larger probability of success \cite{bavarescoStrictHierarchyParallel2021}.
Such strategies cannot be modelled in the tensor network formalism (to our knowledge) and thus, here, we are not interested in anything beyond adaptive strategies.}).

To describe a quantum comb, we use the formalism presented in Ref.~\cite{chiribellaTheoreticalFrameworkQuantum2009}.
Every quantum object is represented by its Choi-Jamiołkowski (CJ) operator.
The CJ operator of a network composed of smaller quantum objects (i.e. composition of two channels, a measurement applied to a state, etc.) is obtained via the link product (denoted by $*$) of the operators of connected elements.
For a matrix representing a linear operator to be a valid quantum comb and for it to respect the causal order between subsequent spaces, it needs to satisfy several trace conditions.

For quantum channels, we will also use the Kraus operators description.
The action of a quantum channel $\mathcal{C}$ from $\mathcal{H}_I$ to $\mathcal{H}_O$ is given by
\begin{equation}\label{eq:krauses}
    \mathcal{C}(\cdot)  = \sum_k K_{k} (\cdot) K_{k}^\dagger
\end{equation}
and Kraus operators $K_k$ obey $\sum_kK_k^\dagger K_k=\id$.
The CJ operator $C \in \mathcal{L}(\mathcal{H}_{O} \otimes \mathcal{H}_{I})$ can be computed from the Kraus operators:
\begin{equation}\label{eq:krauses-to-choi}
    C = \sum_k \ket{K_k}\!\rangle\langle\!\bra{K_k},
\end{equation}
where $\ket{K_k}\!\rangle = (K_k \otimes \id_I) \ket{\id}\!\rangle$
and $\ket{\id}\!\rangle$ is the maximally entangled state on $\mathcal{H}_O \otimes \mathcal{H}_I$.
From now on, when we say that an object is a channel, state or measurement, we mean its CJ operators.

First, we present the description of the channel discrimination task in the language of testers.
This formulation has been proposed in Ref.~\cite{bavarescoStrictHierarchyParallel2021}.

\begin{definition}\label{def:tester}
    A (adaptive) tester for the problem of discrimination between $K$ channels with $N$ uses is a tuple $\{T_k\}_{k=1}^K \in \mathcal{L}(\bigotimes_{n=1}^N \mathcal{H}_{I_n} \otimes \mathcal{H}_{O_n})$,  such that $T_k \geqslant 0$ and $W = \sum_{k=1}^{K} T_k$ satisfies the conditions:
    \begin{equation}\label{eq:tester_cond}
    \begin{split}
        \Tr_{I_{n+1},O_{n+1},\ldots,O_N}[W] &= \Tr_{O_n,I_{n+1},O_{n+1},\ldots,O_N}[W] \otimes \frac{\id_{O_n}}{d_O} \\
        &\qquad\qquad \text{for} \quad n = 1,2,3,...,N,\\
        \Tr [W] &= d_O^n,
    \end{split}
    \end{equation}
where $\mathcal{H}_{I_n}, \mathcal{H}_{O_n}$ denote the input and output spaces of the $n$-th use of the probed channel and $d_I, d_O$ are their dimensions.
The expression $\Tr_S$ denotes the partial trace on subspace $\mathcal{H}_S$ and $\id_S$ the identity operator on $\mathcal{H}_S$.
\end{definition}

The probability we guess $C_k$ correctly is given by
\begin{equation}
    p^{(k)}_{\textrm{succ}} = T_k * C_k^{\otimes N} = \Tr [T_k^{\T} C_k^{\otimes N}].
\end{equation}
And, summing over channels with respective priors, we get the average success probability given by:
\begin{equation}\label{eq:tester_prob}
    p_{\textrm{succ}} =  \sum_k p_k \Tr [T_k^{\T} C_k^{\otimes N}].
\end{equation}

Because of the transpose operation in the Eq.~\eqref{eq:tester_prob} can be omitted (we optimise over the set of testers which is closed under transposition), the task of quantum channel discrimination is now described with the following semidefinite optimisation problem, first presented in Ref.~\cite{chiribellaTheoreticalFrameworkQuantum2009} and in this form in Ref.~\cite{bavarescoStrictHierarchyParallel2021}:
\begin{equation}
\begin{split}
    \textbf{maximise  }\qquad &\sum_k p_k \Tr [T_k C_k^{\otimes N}]\\
    \textbf{subject to}\qquad &\{T_k\} \text{ is a valid tester.}
\end{split}
\end{equation}
Using the conditions from Eq.~\eqref{eq:tester_cond} we translate this problem into:

\begin{equation}\label{eq:SDP}
    \begin{split}
        \textbf{maximise  }\quad &\sum_k p_k \Tr [T_k C_k^{\otimes N}]\\
        \textbf{subject to}\quad &\Tr_{I_{n+1},O_{n+1},\ldots,I_N,O_N}[\sum_kT_k] =\\
        &= \Tr_{O_n,I_{n+1},O_{n+1},\ldots,I_N,O_N}[\sum_kT_k] \otimes \frac{\id_{O_n}}{d_O} \\
        &\qquad\qquad\qquad\qquad \text{for} \quad n = 1,2,3,...,N,\\
        &\Tr \left[\sum_kT_k\right] = d_O^N, \quad T_k \geqslant 0 \text{ for all }k.
\end{split}
\end{equation}
This problem may be efficiently solved using publicly available SDP solvers, provided the dimensions of the relevant Hilbert spaces and number of channel uses is small enough. For our numerical computations, we have employed the CVXPY package \cite{diamondCVXPYPythonEmbeddedModeling2016} and the MOSEK solver \cite{MOSEK2024}.

\begin{figure*}[t]
    \centering
    \includegraphics[scale=0.8]{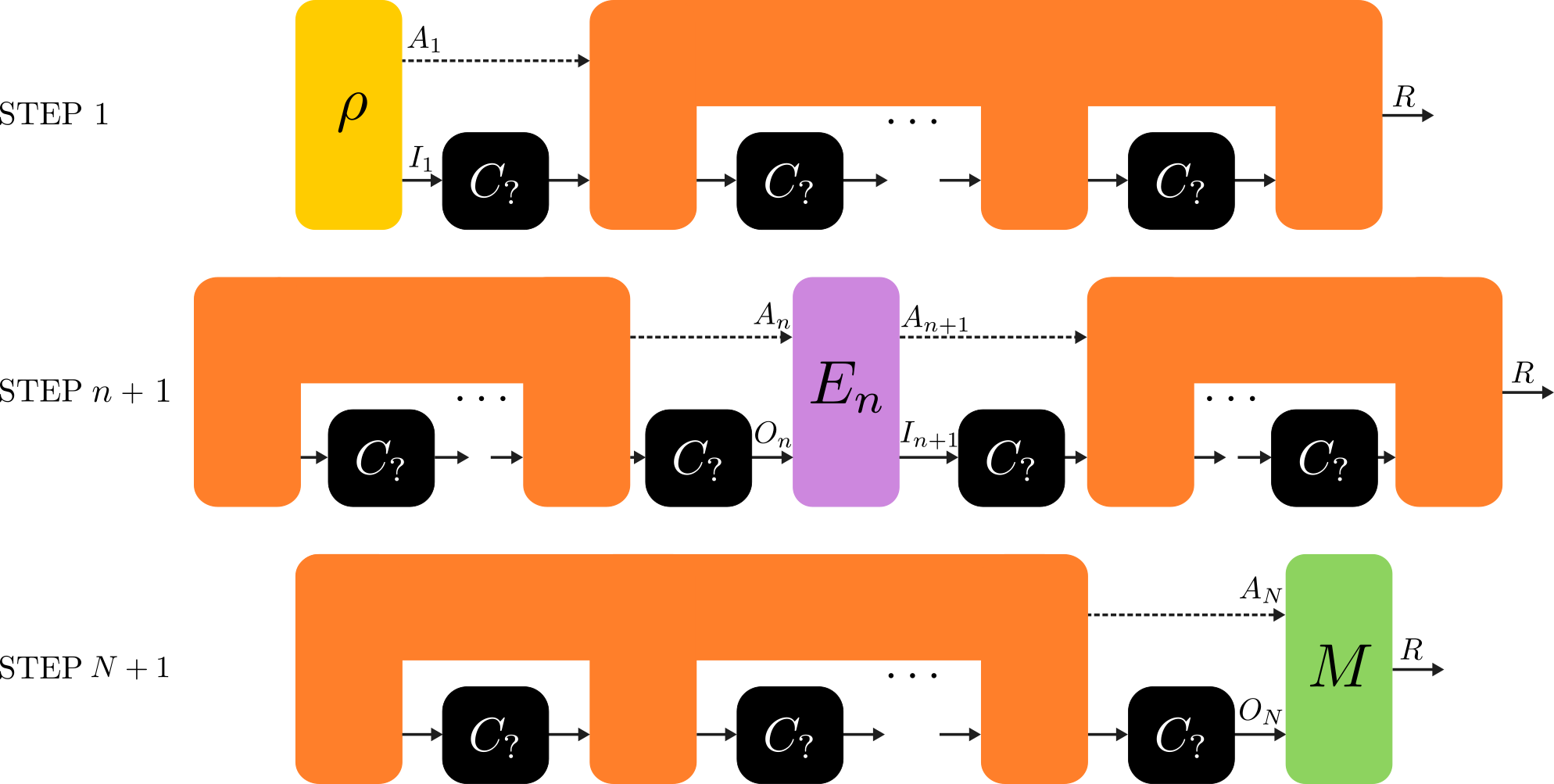} 
    \caption{Diagrammatic representation of our tensor network algorithm.
    The boxes represent the unknown channel $C_?$ and the parts of the discrimination strategy.
    The arrows represent input, output and ancilla spaces, some of them labelled as in Def.~\ref{def:strategy}.
    The orange shapes represent combined parts of the comb that remain fixed during one optimisation step.
    In every step, the whole comb outside the currently optimised tooth is treated as a constant.
    The repeated optimisation over the teeth (input state $\rho$, inter-channel quantum controls $E_n$ and `measurement channel' $M$) hopefully leads to an optimal comb.}
    \label{fig:iterations}
\end{figure*}

\section{Tensor network algorithm}
\label{sec:algorithm}

The main problem with implementation of the algorithm presented in the previous section, is the usual one that haunts all quantum related research: when we have more objects (more qubits, more gates, more channels etc.) we face exponentially growing dimensions of the Hilbert spaces we need to deal with.
The SDP will be efficient when, e.g. discriminating qubit-qubit channels with 3-4 channel uses, but with growing problem size the size of the corresponding tester matrices will scale exponentially with the number of channel uses.

If we want to obtain insight into the regime of large number of channel-uses (ideally in order to understand what happens in the asymptotic limit $N\rightarrow \infty$) we need to follow another path.
We can redefine the discrimination strategy, by coming back from the full comb description to the input state, measurement, and quantum control operations we are performing between channel uses.
\begin{definition}\label{def:strategy}
    An adaptive measurement strategy for the problem of discrimination between $K$ channels with $N$ uses consists of an input state $\rho \in \mathcal{L}(\mathcal{H}_{I_1} \otimes \mathcal{H}_{A_1})$, inter-channel quantum controls $E_n \in \mathcal{L}(\mathcal{H}_{A_{n+1}} \otimes \mathcal{H}_{I_{n+1}} \otimes \mathcal{H}_{A_n} \otimes \mathcal{H}_{O_n})$ and a `measurement channel' $M\in \mathcal{L}(\mathcal{H}_{A_N} \otimes \mathcal{H}_{O_N} \otimes \mathcal{H}_R)$, which obey the following trace conditions
    \begin{equation}\label{eq:strategy-cond}
    \begin{split}
        \Tr_{I_1,A_1}\rho&=1\\
        \Tr_{I_{n+1},A_{n+1}} E_n &= \id_{O_n,A_n}\\
        \Tr_{R}M &= \id_{O_N,A_N}.
    \end{split}
    \end{equation}
\end{definition}

We formally model the final measurement step as a `measurement channel' with $K$-dimensional output space representing the label of the channel that is guessed.
The channel is guessed by computational basis measurement on the space $\mathcal{H}_R$.
The full strategy can be represented by a comb $\tilde{W}$, which is a network combining all its elements:
\begin{equation}\label{eq:tester_tensor}
    \tilde{W} = M * E_{N-1} * \ldots * E_1 * \rho,
\end{equation}

In this new language, the probability of success is given by
\begin{equation}\label{eq:probability-tensor}
    p_{\t{succ}} = \sum_k p_k \tilde{W} * C_k^{\otimes N}*P_k = \sum_k p_k \langle k|\rho_k |k\rangle,
\end{equation}
where $P_k$ is the CJ operator of the $|k\rangle\langle k|$ projector on space $\mathcal{H}_R$ and $\rho_k$ is the output state on space $\mathcal{H}_R$ when the strategy is used with channel $C_k$.

The diagram in Fig. \ref{fig:iterations} shows the tensor network decomposition of a comb and the `teeth' which are to be optimised over.
Instead of the memory-costly optimisation over the whole comb, we can optimise over the input state $\rho$, inner transformations -- `teeth' $E_n$ and measurement $M$ independently. In this way, provided we can restrict ourselves to small dimensional ancillary spaces $A_n$, we may perform the optimization also in the large $N$ limit. 
This is the essence of the tensor-network approach we propose here.
The optimisation runs as follows.

We initialise $\rho$, $E_n$ with random values that satisfy the trace conditions from Eq.~\eqref{eq:strategy-cond} 
In a single step of optimisation, we compute the link product of all other teeth and the $N$ channel uses.
Then we create an SDP to maximise the link product of the optimised tooth and the rest of the comb ($\Tr [\text{comb}^{\T} \text{tooth}]$), with constraints from Eq.~\eqref{eq:strategy-cond} and treating the rest of the comb as a constant.
Operationally, in each step we maximise the value
\begin{equation}\label{eq:probability-algorithm-step}
    \sum_k p_k \tilde{W}_{\t{right}} * C_k^{\otimes (N-n)} * E_n * \tilde{W}_{\t{left}} * C_k^{\otimes n},
\end{equation}
with $\rho$ for $n=0$ and $M$ for $n=N$, with $\tilde{W}_{\t{right}}, \tilde{W}_{\t{left}}$ the parts of the comb after and before the currently optimised tooth.
We subsequently optimise for $\rho, E_1, \ldots, E_{N-1}, M$ and then from $\rho$ again in a cycle.
We stop the optimisation after the success probability does not show relative improvement larger than some fixed $\varepsilon$ over a specified number of cycles (we typically choose five).
Value of $\varepsilon$ is one of the parameters of the algorithm. In order to obtain a relatively fast algorithm we typically chose $\varepsilon = 10^{-4}$ but in some cases choosing lower $\varepsilon$ helped achieve larger success probabilities.

Until now, we have not paid much attention to the ancillary spaces.
For the general SDP solution, the ancilla exists `somewhere in the comb' implicitly.
Now, because we treat each tooth individually, we need to make a decision on the dimension of ancilla we want to implement.
Smaller ancilla is cheaper to compute (the teeth matrices are smaller and thus SDPs use less memory and are computed faster), but, for some channels, discrimination needs larger ancilla (as we argue later, possibly a lot larger).

Before moving on to the results of this numerical approach, we will first discuss some intricate relations between quantum discrimination and estimation problems. This will allow us to obtain a deeper insight into the character of the results we will be getting as a result of numerical optimisation of optimal discrimination strategies, as well as provide tools to compute bounds on discrimination problems that arise from quantum metrological analysis. 

\section{Channel estimation perspective on the channel discrimination problem}
\label{sec:estimation}

The tasks of channel discrimination and estimation are closely related.
The objective of quantum channel estimation is to estimate a parameter (or multiple parameters) of a channel that belongs to some parametrized family of quantum channels.
Discrimination can be thought of as a particular subclass of estimation in which we assume that the parameter has a discrete set of values, unlike in typical estimation problems where it is continuous.
Still, given a discrimination problem, one may consider a corresponding estimation problem, where parameters are being continuously interpolated between the discrete values. In this way, as we show below, it is possible to upper bound the channel discrimination probability of success using an expression involving the QFI for the estimation problem, integrated over the continuous trajectory connecting the discrete parameter values.  

\subsection{Discrimination bounds based on QFI}
\label{subsec:bounds}
Let us focus on a two-channel discrimination problem.
Definitions of different distances between states used here, the inequalities between them and their connection to QFI are described in Ref.~\cite{albarelliProbeIncompatibilityMultiparameter2022}.
If we think of the state of the system (and ancilla) before the measurement, we can derive the probability of success from the trace distance of the states generated by applying different channels.
Let $\rho_1,\rho_2$ be the states before measurement when channels $C_1,C_2$ respectively are given for the discrimination strategy.
Then, the maximal probability of success is given by the famous Helstrom formula \cite{helstromQuantumDetectionEstimation1969}:
\begin{equation}\label{eq:trace}
    p_{\textrm{succ}} = \frac{1}{2} \left(1+D_{\textrm{tr}}(\rho_1,\rho_2)\right),
\end{equation}
where $D_{\textrm{tr}}$ is the trace distance.
Or equivalently, the minimal error probability:
\begin{equation}\label{eq:helstrom-error}
    p_{\textrm{err}} = \frac{1}{2} \left(1-D_{\textrm{tr}}(\rho_1,\rho_2)\right).
\end{equation}
Trace distance is then related to the Bures' angle \cite{Bengtsson2006} via
\begin{equation}\label{eq:bures_trace}
    D_{\textrm{tr}}(\rho_1,\rho_2) \leqslant \sin D_\mathrm{A}(\rho_1,\rho_2).
\end{equation}
The infinitesimal Bures' angle on the other hand is connected to QFI:
\begin{equation}\label{eq:bures}
    D_\mathrm{A} (\rho_\theta, \rho_{\theta + \mathrm{d}\theta}) = \frac{1}{2}\sqrt{\mathcal{F}(\rho_\theta)}\mathrm{d}\theta.
\end{equation}
And thus the angle is bounded by an integral
\begin{equation}\label{eq:bures_bound}
    D_\mathrm{A} (\rho_{\theta_1}, \rho_{\theta_2}) \leqslant \frac{1}{2}\int_{\theta_1}^{\theta_2}\sqrt{\mathcal{F}(\rho_\theta)}\mathrm{d}\theta.
\end{equation}
This connection was also made in \cite[see section III and Eq. (26) particularly]{yuanFidelityFisherInformation2017} which was brought to our attention after the completion of the first draft.

If we now consider a unitary parameter estimation model where 
$\rho_\theta = U_\theta \rho U_\theta^\dagger$, $U_\theta=\exp(\I G \theta)$, where $G$ is some Hermitian generator, then the QFI will not depend on the value of $\theta$ \cite{Giovaennetti2006, zhouAsymptoticTheoryQuantum2021}.
If we now consider a channel discrimination problem, where the corresponding discrete parameter values are set to be:
$\theta_1 = 0, \theta_2 = \Delta \theta$, from \eqref{eq:bures_bound} we get
\begin{equation}\label{eq:integrated}
    D_\mathrm{A} (\rho_0, \rho_{\Delta\theta}) \leqslant \frac{\Delta \theta}{2}\sqrt{\mathcal{F}},
\end{equation}
and combining \eqref{eq:trace}, \eqref{eq:bures_trace} and \eqref{eq:integrated}, we arrive at
\begin{equation}\label{eq:final-bound-single}
    p_{\textrm{err}} \geqslant \frac{1}{2} \left[ 1 - \sin\left( \frac{\Delta \theta}{2}\sqrt{\mathcal{F}}\right) \right].
\end{equation}

The real benefit of the above bound, may be appreciated, if we note that we can apply it as well to lower bound the error probability of the most general $N$-channel use discrimination strategy, 
\begin{equation}\label{eq:final-bound-n}
    p^{(N)}_{\textrm{err}} \geqslant \frac{1}{2} \left[ 1 - \sin\left( \frac{\Delta \theta}{2}\sqrt{\mathcal{F}_N} \right) \right].
\end{equation}
where $\mathcal{F}_N$ is the maximal achievable QFI for estimation of $\theta$ in the most general $N$-channel use adaptive strategy (assuming again $\mathcal{F}_N$ does not depend on $\theta$). 
This allows us now to apply powerful and efficiently computable bounds on $\mathcal{F}_N$ developed in the field of quantum metrology \cite{escherGeneralFrameworkEstimating2011, demkowicz-dobrzanskiElusiveHeisenbergLimit2012a, demkowicz-dobrzanskiUsingEntanglementNoise2014, zhouAsymptoticTheoryQuantum2021, kurdzialekUsingAdaptivenessCausal2023}

Let $\{K_{k,\theta}\}_k$ be the Kraus decomposition \ref{eq:krauses} of the channel we want to estimate the parameter of.
Given $N$ uses of the channel, one can obtain an upper bound on the QFI achievable after $N$ channel uses $\mathcal{F}_N$, with the most general adaptive estimation strategy via the following iterative algorithm \cite{kurdzialekUsingAdaptivenessCausal2023}:
\begin{equation}\label{eq:qfi_bound}
    \mathcal{F}_{N+1} = \mathcal{F}_N + 4 \min_{h} \left(\|\alpha(h)\| + \sqrt{\mathcal{F}_N}\|\beta(h)\|\right),
\end{equation}
where the minimisation is over Hermitian matrices $h$, which size corresponds to the number of Kraus operators in the chosen Kraus representation of the channel, $\|\cdot\|$ is the operator norm and $\alpha(h),\beta(h)$ are defined as:
\begin{multline}\label{eq:alfa-beta}
    \alpha(h) = \sum_i \Big(\dot{K}_i - i \sum_j h_{ij} K_j\Big)^\dagger \Big(\dot{K}_i - i \sum_j h_{ij} K_j\Big),\\
    \beta(h) = i \sum_i K_i^\dagger \Big(\dot{K}_i - i \sum_j h_{ij} K_j\Big),
\end{multline}
where $\dot{K}_i$ represents the parameter derivative of a Kraus operator, and we have dropped the explicit dependence on $\theta$ for compactness.
The iteration starts, with single channel QFI computed via:
\begin{equation}\label{eq:qfi_bound_one}
    \mathcal{F}_1 = 4 \min_{h} \|\alpha(h)\|.
\end{equation}
This iterative bound can be computed in a straight-forward way using the QMetro++ package \cite{dulianQMetroPythonOptimization2025}.

\subsection{Heisenberg scaling estimation and perfect channel discrimination}
\label{subsec:intuition}
An important \emph{qualitative} question in quantum metrology, is whether  a given parameter estimation problem admits the Heisenberg scaling of QFI or not, i.e. whether the QFI scales quadratically with number of uses $N$ or only linearly \cite{Giovaennetti2006, demkowicz-dobrzanskiElusiveHeisenbergLimit2012a}. 
A model that admits Heisenberg scaling may be regarded as `easy' to estimate compared to non-Heisenberg scaling ones.

Using the iterative bound \eqref{eq:qfi_bound} it is clear that if there exist a matrix $h$ such that $\beta(h)=0$ then the obtained bound $\mathcal{F}_N$ will scale at most linearly, and Heisenberg scaling will be impossible \cite{escherGeneralFrameworkEstimating2011, demkowicz-dobrzanskiElusiveHeisenbergLimit2012a, demkowicz-dobrzanskiUsingEntanglementNoise2014}. This is the case if and only if the following  simple algebraic condition, Hamiltonian not in Kraus span (HNKS), is satisfied \cite{zhouAsymptoticTheoryQuantum2021}: 
\begin{equation}
    \label{eq:hnks}
    H = i \sum_i K_i^\dagger \dot{K}_i \notin \mathcal{S}, \quad \mathcal{S} = \text{span}_\mathbb{H}\{K_i^\dagger K_j\}_{i j}.
\end{equation}
where $H$ is the effective `parameter encoding Hamiltonian' of the problem.

In case of a channel discrimination problems, an interesting qualitative difference is whether the channels can be discriminated perfectly with finite number of channel uses or not. 
According to \cite{duanPerfectDistinguishabilityQuantum2009} a pair of channels $C_0, C_1$ with Kraus operators $\{K_{0i}\}, \{K_{1i}\}$ can be perfectly discriminated by finite number of uses if and only if (i): the channels are disjoint (there exists an input state such that there is no non-trivial intersection between supports of the output density matrices) and (ii):
\begin{equation}
    \label{eq:finite-discrimination}
    \id \notin \text{span}_{\mathbb{H}}\{K_{1i}^\dagger K_{2j}\}_{i j}.
\end{equation} 
Focusing on the second condition, it shows some apparent similarity with the Heisenberg scaling condition \eqref{eq:hnks}. 
Below, we provide an  argument indicating that there is indeed a 
connection between the finite-use distinguishability in a channel discrimination model and the Heisenberg scaling for estimation of a local continuous counter-part of the same model.

Let us now assume the Hamiltonian of our channel estimation model $C_\theta$ at $\theta=0$ \emph{does not satisfy} the HNKS condition, i.e. the model does not admit Heisenberg limit of precision.
Let $K_i$ and $\dot{K}_i$ be the Kraus operators of the model and their derivatives taken at $\theta=0$.
Let us now consider a channel discrimination problem, 
between channels $C_1=C_{\theta=0}$ and $C_2 \approx C_{\theta=\Delta \theta}$ for which the respective Kraus operators are defined as:
     \begin{align}
    \label{eq:discrimination_kraus}
    K_{1,i} &= K_i,  
    &K_{2,i} &= \sqrt{1-\varepsilon}(K_i + \Delta\theta\dot{K}_i), \\
    & &K_{2,0} &= \sqrt{\id-\sum_{i=1} K_{2,i}^\dagger K_{2,i}},
    \end{align}
where $\varepsilon \geqslant 0$ is some free parameter to be set later.
Note that $C_2$ does not exactly correspond to $C_{\theta=\Delta \theta}$, since we have truncated the Taylor expansion of Kraus operators at the first order. As a result, we also needed to include 
the additional Kraus operator $K_{2,0}$ to ensure that trace preservation condition is strictly obeyed despite truncating the Taylor series, $K_{2,0}^\dagger K_{2,0} + \sum_k K_{2,i}^\dagger K_{2,i} = \id$.

Let us now discuss the choice of parameter $\varepsilon$. 
Ideally, we would like to set $\varepsilon \approx 0$, to make
$C_2$ as similar to $C_{\theta=\Delta \theta}$ as possible for small $\Delta \theta$.  We need to make sure, though, that 
in the definition of $K_{2,0}$ in \eqref{eq:discrimination_kraus} the operator under the square-root in is positive, as otherwise this construction would not make sense. 
For this to be the case we need to satisfy:
\begin{multline}
    \label{eq:positive-kraus}
    \id-\sum_{i=1} K_{2,i}^\dagger K_{2,i} = \id - (1-\varepsilon)\Big(\sum_{i=1} K_i^\dagger K_i + \Delta\theta\dot{K}_i^\dagger K_i +\\
    + \Delta\theta K_i^\dagger \dot{K}_i + \Delta\theta^2\dot{K}_i^\dagger \dot{K}_i\Big) =\\
    = \id - (1-\varepsilon)\left(\id + \sum_{i=1}\Delta\theta^2\dot{K}_i^\dagger \dot{K}_i\right) \overset{!}{\geqslant} 0.
\end{multline}
This condition is equivalent to
\begin{equation}
    \label{eq:positive-kraus-contd}
    \varepsilon\id - (1-\varepsilon)\sum_{i=1}\Delta\theta^2\dot{K}_i^\dagger \dot{K}_i \overset{!}{\geqslant} 0,
\end{equation}
which becomes true for $\epsilon = a \Delta\theta^2$ for some big enough $a$. This implies, that if we want to consider discrimination of channels corresponding to small parameter shift $\Delta \theta$, the required value of $\varepsilon$ will go to zero quadratically faster 
$\varepsilon \propto \Delta \theta^2$. 

Now let us compute the following expression involving Kraus operators of the two channels:
\begin{multline}
    \label{eq:hnks-to-finite-discrimination}
    \sum_i K_{1,i}^\dagger K_{2,i} = \sum_i \sqrt{1-\varepsilon}K_i^\dagger (K_i + \Delta\theta\dot{K}_i) = \\
    = \sqrt{1-\varepsilon}\id + \sqrt{1-\varepsilon}\Delta\theta\sum_i K_i^\dagger \dot{K}_i.
\end{multline}

Since we have assumed that $C_{\theta}$ does not satisfy the HNKS condition \eqref{eq:hnks}, this implies that 
$H$ may be written as  $H = \sum_{i,j} \tilde{h}_{i j} K_i^\dagger K_j$, with $\tilde{h}$ being some Hermitian matrix.
Thus,
\begin{multline}
    \label{eq:hnks-to-finite-discrimination-2}
    \sum_i K_{1,i}^\dagger K_{2,i} = \sqrt{1-\varepsilon}\left( \id - i \Delta\theta\sum_{i,j} \tilde{h}_{i j} K_i^\dagger \dot{K}_i\right)  
\end{multline}
Consequently:
\begin{equation}
\id = \sum_{i,j} (\delta_{i j} + i\Delta\theta \tilde{h}_{i j}) K_{1,i}^\dagger K_{2,j} + \mathcal{O}(\Delta\theta^2),
\end{equation}
where we used have used the fact that $\varepsilon \propto \Delta \theta^2$, and so $1/\sqrt{1-\varepsilon} \propto 1 + O(\Delta\theta^2)$.  
If, now, the angle $\Delta\theta$ is sufficiently small for the second order terms to be negligible, the above condition clearly violates  the  the finite-channel use discrimination condition \eqref{eq:finite-discrimination}! 

We can, therefore, conclude that given a channel estimation model which \emph{does not allow} for the Heisenberg scaling of precision, 
the corresponding discrimination model, obtained channels differing by a small value of parameter $\Delta \theta$, will not allow for perfect discrimination with finite number of channel uses. Clearly, the above statement is not very operational, as it rigorously holds only in the limit $\Delta \theta \rightarrow 0$, for which the discrimination problem is ill-defined. 

Nevertheless, we may treat this result as a guideline, telling us what to expect, whenever we consider discrimination models with smaller and smaller $\Delta \theta$,  for which the corresponding estimation models manifest Heisenberg scaling or not. We will see in the next section, that indeed the qualitative relation we have highlighted above nicely explains the character of the dependence of discrimination success probability with increasing number of channel uses, depending on the character of the corresponding estimation model.  

Additionally,  we may also make use of the bound \eqref{eq:final-bound-single}, and observe that for small $\Delta \theta$
\begin{equation}
\label{eq:prberrexpectedscaling}
 p^{(N)}_{\textrm{err}} \overset{\Delta \theta \rightarrow 0}{\geqslant} \frac{1}{2} - \begin{cases}{ \t{const} \times N} & \t{Heisenberg scaling}  \\
 \t{const} \times \sqrt{N} &\t{non-Heisneberg scaling.} 
 \end{cases}
\end{equation}
Hence, we may expect that if the bound is tight enough (which is expected in the small $N$ regime) the initial character drop of discrimination error will be either a fast linear drop or slow square root drop. We cannot base any large $N$ prediction based on this bound, however, as it will become less and less tight for increasing $N$. In particular, when $\sqrt{\mathcal{F}_N} \Delta \theta/2 = \pi/2$, the bound, $p_{\t{err}}^{(N)} \geq 0$, becomes trivial, and from this point does not provide us with any useful information.

\section{Examples}
\label{sec:examples}

In this section, we illustrate the numerical efficiency of the method discussed in Sec.~\ref{sec:algorithm} for a number of 
important channel discrimination problems, comparing them with the quantum metrology bounds as well as highlighting the qualitative
connection between Heisenberg scaling and finite-channel use discrimination discussed in previous section.

When performing numerical search for the optimal discrimination protocols, we will also gradually increase the dimension of the ancillary system $d_A$ involved in order to assess whether entanglement with ancillary systems is needed to reach the optimal performance of the discrimination protocol.

\subsection{Unitary channels}
\label{subsec:unitaries}
As a sanity check of our numerical method, we first consider, the basic problem of discrimination of two
unitary channels, for which the exact solution is known.
The minimal error probability is given by \cite{acinStatisticalDistinguishabilityUnitary2001}:
\begin{equation}\label{eq:unitary_psucc}
    p_{\textrm{err}} = \frac{1}{2}\left[1 - \sin (N\alpha)\right],
\end{equation}
where $N$ is the number of uses and 
\begin{equation}\label{eq:alpha-angle}
    \alpha = \arccos \left(\frac{| \Tr(U_1^\dagger U_2)|}{2}\right)
\end{equation}
is the separation angle between the channel matrices.
Let us consider qubit unitary channels, where without loss of generality, we may consider 
the unitaries (which are at the same time the single Kraus operators of the channels)
\begin{equation}
    U_1=\id \quad \textrm{and} \quad U_2 = e^{-\frac{i}{2}\Delta\theta \sigma_z} \quad \textrm{where} \quad \sigma_z = \left[
    \begin{smallmatrix}
        1 & 0\\
        0 & -1
    \end{smallmatrix}
    \right]
    .
\end{equation}
For this model the angle $\alpha$ is
\begin{equation}
    \cos^2 \alpha = \frac{1}{4}\left|\Tr\left[
        \begin{smallmatrix}
            e^{-\frac{i}{2}\Delta\theta} & 0\\
            0 & e^{\frac{i}{2}\Delta\theta}
        \end{smallmatrix}
        \right]\right|^2 = \cos^2 \frac{\Delta\theta}{2} \Rightarrow \alpha = \frac{\Delta\theta}{2}
\end{equation}

In Fig.~\ref{fig:perfect_discr} (upper) we show the result of our optimization which confirm that analytical bound, and show that 
even without the use of any ancillary system this ideal probability is attained by our optimisation
(footnote \footnote{One optimal strategy is to choose the inter-channel teeth to be the inverse of one of the unitaries, thus getting a one use discrimination between identity and a unitary -- the product of the first one and inverse of the second. Then the input state is chosen perpendicular to the rotation axis of the unitary on the Bloch sphere.}).

Clearly, the corresponding estimation model $U_{\Delta\theta} = e^{-\frac{i}{2}\Delta\theta \sigma_z} $ is a noise-less phase estimation model, for which the Heisenberg scaling is achievable $\mathcal{F}_N \propto N^2$. This is in agreement with our qualitative intuition of Sec.~\ref{subsec:intuition} that finite-channel use discrimination should only be possible if the underlying model manifest Heisenberg scaling. 
Moreover, for small $N$, the error probability decreases linearly as expected from \eqref{eq:prberrexpectedscaling}.

\subsection{Perpendicular dephasing---signal-first order}
\label{subsec:perp-trivial}

We now move on to consider a more challenging problem, of discrimination of unitary rotations with additional noise added in the direction perpendicular to this rotation.
The Kraus operators for the two channels $C_1= C_{\theta=0}$, $C_2 = C_{\theta=\Delta\theta}$, where $C_\theta$ is the channel with the following Kraus operators:
    \begin{align}
    \label{eq:perp-deph_trivial}
    K_{\theta,1} &= \sqrt{p}e^{-\frac{i}{2}\theta \sigma_z},&  K_{\theta,2} &= \sqrt{1-p} \sigma_xe^{-\frac{i}{2}\theta \sigma_z}.
    \end{align}
In the above model the unitary $\Delta\theta$ rotation comes first (the signal), before the noise. 
It is known \cite{demkowicz-dobrzanskiAdaptiveQuantumMetrology2017, zhouAchievingHeisenbergLimit2018, arradIncreasingSensingResolution2014, kesslerQuantumErrorCorrection2014, durImprovedQuantumMetrology2014, sekatskiQuantumMetrologyFull2017} that for this channel estimation problem, we can effectively remove noise through an error-correction procedure, and obtain and recover the ideal Heisenberg scaling of QFI.
Moreover, thanks to the fact that the signal comes before the noise, the quantum error correction procedure is independent of the value of $\theta$. 

\begin{figure}[t!]
    \centering
    \includegraphics[scale=0.25]{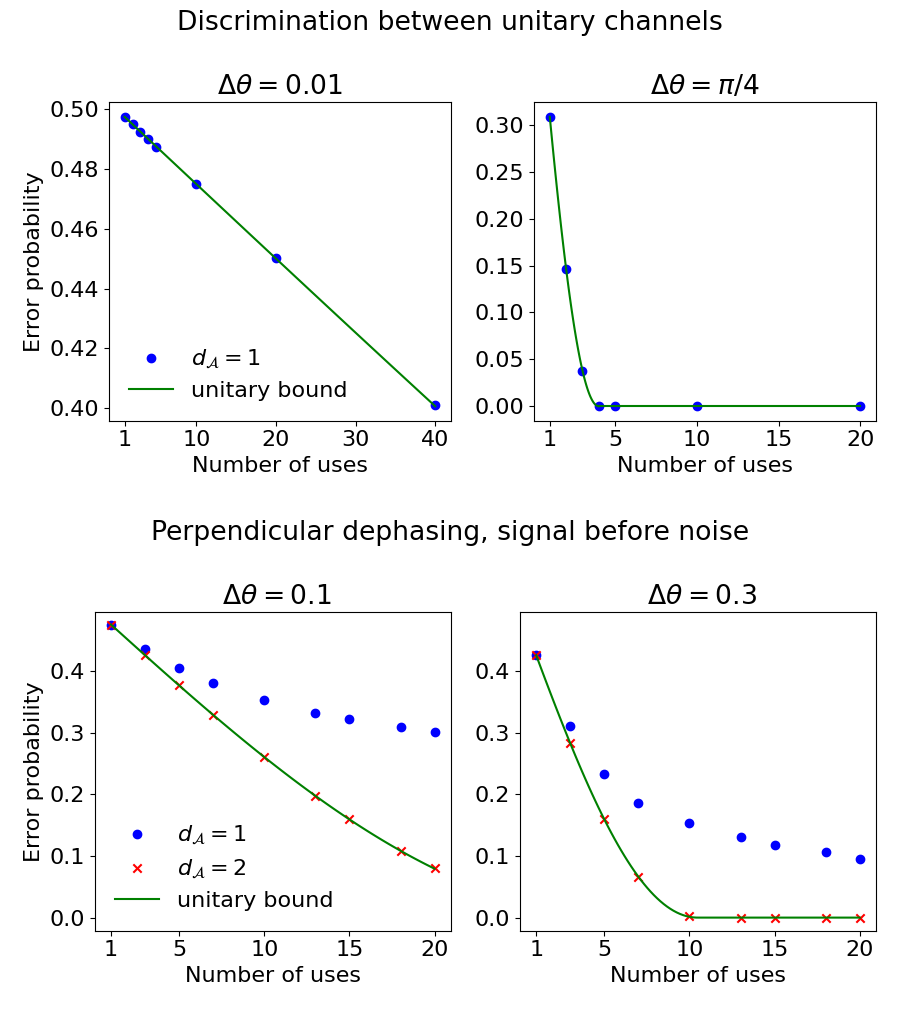}
    \caption{Error probability of discrimination between two unitaries (upper) and unitaries with perpendicular dephasing noise (lower) for different number of channel uses and ancilla dimensions 1 and 2. The solid green line represents the theoretical probability of error for unitary discrimination---a lower bound for the numerical results. The effect of the noise can be got rid of with a single qubit ancilla.}
    \label{fig:perfect_discr}
\end{figure}

In the lower plots of Fig.~\ref{fig:perfect_discr} we 
plot the reduction of discrimination error probability with increasing number of channel uses, for two different channel separations $\Delta\theta=0.1$ and $\Delta\theta =0.3$. We see that, in case no ancillary system us 
is used ($d_A=1$), we are not able to perform perfect discrimination. 

If, however, we have a single qubit ancilla ($d_A=2$), we already recover the optimal discrimination procedure that yields error-probability that coincides with the ideal unitary discrimination problem and allows for perfect discrimination in finite number of uses.
This comes as no surprise, as the quantum error-correction procedure that is known from metrological considerations \cite{demkowicz-dobrzanskiAdaptiveQuantumMetrology2017, zhouAchievingHeisenbergLimit2018, arradIncreasingSensingResolution2014, kesslerQuantumErrorCorrection2014, durImprovedQuantumMetrology2014, sekatskiQuantumMetrologyFull2017} also requires only a single qubit ancillary system.

\subsection{Perpendicular dephasing---noise-first order}
\label{subsec:perp-after}
The Kraus operators for this noise channel are
\begin{equation}\label{eq:perp-deph}
    K_{\theta,1} = \sqrt{p}e^{-\frac{i}{2}\theta \sigma_z}, \quad K_{\theta,2} = \sqrt{1-p} e^{-\frac{i}{2}\theta \sigma_z}\sigma_x
\end{equation}
and so the only difference, compared with \eqref{eq:perp-deph_trivial}, is that the unitary rotation comes \emph{after} the noise. 
From a quantum metrological perspective, this model is similar to the previous one, as it also admits Heisenberg scaling of the optimal QFI. 
Nevertheless, the actual quantum error-correction protocol differs, with varying value of parameter $\theta$, and hence there is no single parameter-independent quantum-error correction protocol that would work for all $\theta$. We may expect that this will cause some complications, when we try to discriminate, channels that are at finite separation $\Delta \theta$. 

\begin{figure}[t!]
    \centering
    \includegraphics[scale=0.25]{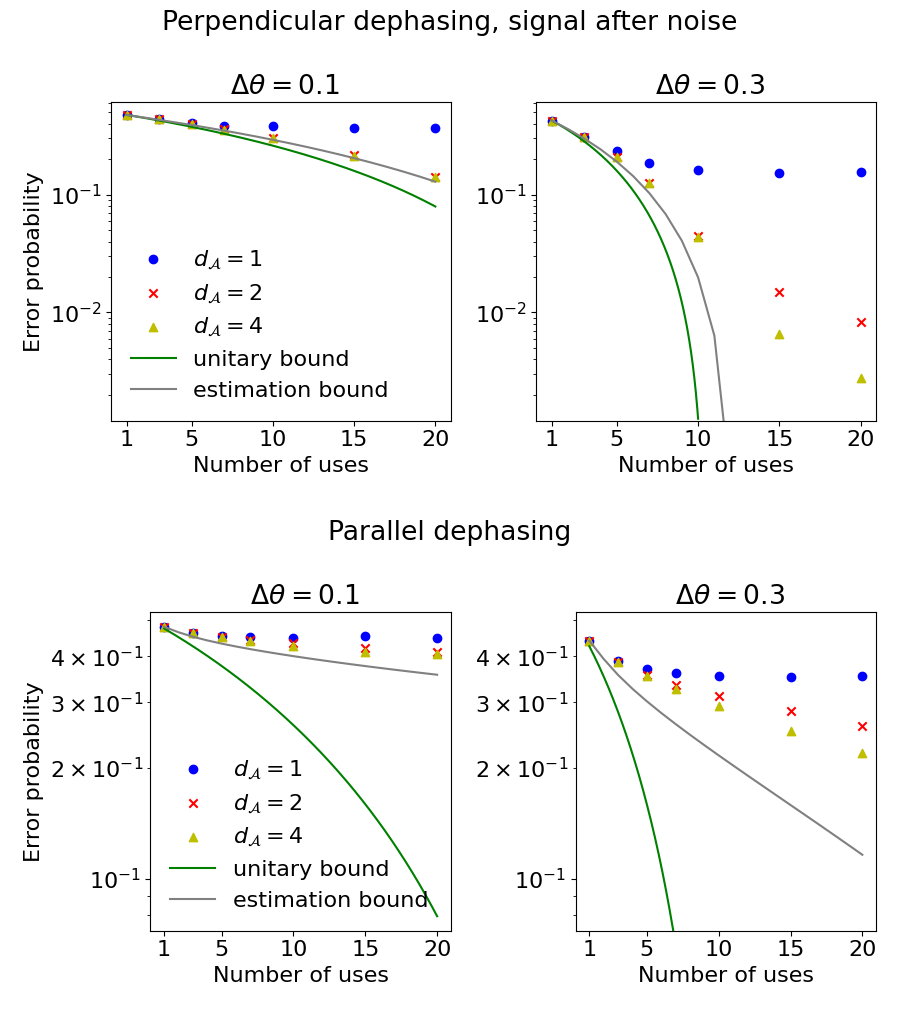}
    \caption{Error probability of discrimination between two unitaries with perpendicular (upper) and parallel (lower) dephasing noise for different number of channel uses and ancilla dimensions 1, 2 and 4. The grey solid lines show the estimation bounds computed with the QMetro++ package, and the green lines are the unitary bound. Logarithmic scale in y-axis.}
    \label{fig:bounds}
\end{figure}

The numerical results of our algorithm for the noise-first order perpendicular dephasing model can be seen in the upper plot of Fig.~\ref{fig:bounds}.
For the small angle regime, the numerical results come close to attaining the estimation bound \eqref{eq:final-bound-n}.
We also see that initial drop in probability is linear, as predicted by \eqref{eq:prberrexpectedscaling}.
However, even though the initial behaviour might indicate that the error will eventually drop to zero for a finite number of uses, we see for larger separations $\Delta\theta=0.3$, that this is not the case. We should attribute this to the fact, that for these large separations we are not able to perform quantum error-correction such that both channels are effectively mapped into unitaries.  

Because the model is not perfectly distinguished by finite uses, asymptotically the error probability should scale exponentially, see \cite{nuradhaQueryComplexityClassical2025a}.

\subsection{Parallel dephasing}
\label{subsec:parallel}
The Kraus operators for this noise channel are
\begin{equation}\label{eq:par-deph}
    K_{\theta,1} = \sqrt{p}e^{-\frac{i}{2}\theta \sigma_z}, \quad K_{\theta,2} = \sqrt{1-p} e^{-\frac{i}{2}\theta \sigma_z}\sigma_z.
\end{equation}
This model is studied in \cite{escherGeneralFrameworkEstimating2011,demkowicz-dobrzanskiElusiveHeisenbergLimit2012a,demkowicz-dobrzanskiUsingEntanglementNoise2014,zhouAsymptoticTheoryQuantum2021}, where analytical bounds for the QFI are presented, which prove that 
this model \emph{does not admit} Heisenberg scaling. We should therefore expect worse discrimination performance as well, and do not count on finite channel-use distinguishability. 

The numerical results of our algorithm for the parallel dephasing model can be seen in the upper plot in Fig. \ref{fig:bounds}.
For the small angle regime, the numerical results come close to the QFI bound \eqref{eq:final-bound-n}.
This shows not only that the strategies found by the optimisation are close to optimal, but also that the bound coming from estimation describes well the dependence on number of uses.
For the large angle regime, the bound is less tight, but still provides much better insight into the problem than the unitary discrimination bound. We also see slower than linear initial drop in error probability in agreement with \eqref{eq:prberrexpectedscaling}.

\subsection{Dephasing vs amplitude damping}
\label{subsec:bigancilla}
Finally, moving away from the class of unitary parameter encoded channels, let us investigate a discrimination problem where the goal is to discriminate between two channels representing qualitatively different noise processes: dephasing and amplitude damping.
The Kraus operators of the respective channels are:
\begin{equation}\label{eq:deph_ampl_damp}
    \begin{split}
    K_{1,1} &= \sqrt{\eta}\id, \quad K_{1,2} = \sqrt{1-\eta} \sigma_x\\
    K_{2,1} &= \left[\begin{smallmatrix}
        1 & 0 \\
        0 & \sqrt{1-\gamma}
    \end{smallmatrix}\right], \quad K_{2,2} = \left[\begin{smallmatrix}
        0 & \sqrt{\gamma} \\
        0 & 0
    \end{smallmatrix}\right],
    \end{split}
\end{equation}
where $\eta,\gamma$ represents noise strength parameters.

\begin{figure}[t!]
    \centering
    \includegraphics[scale=0.33]{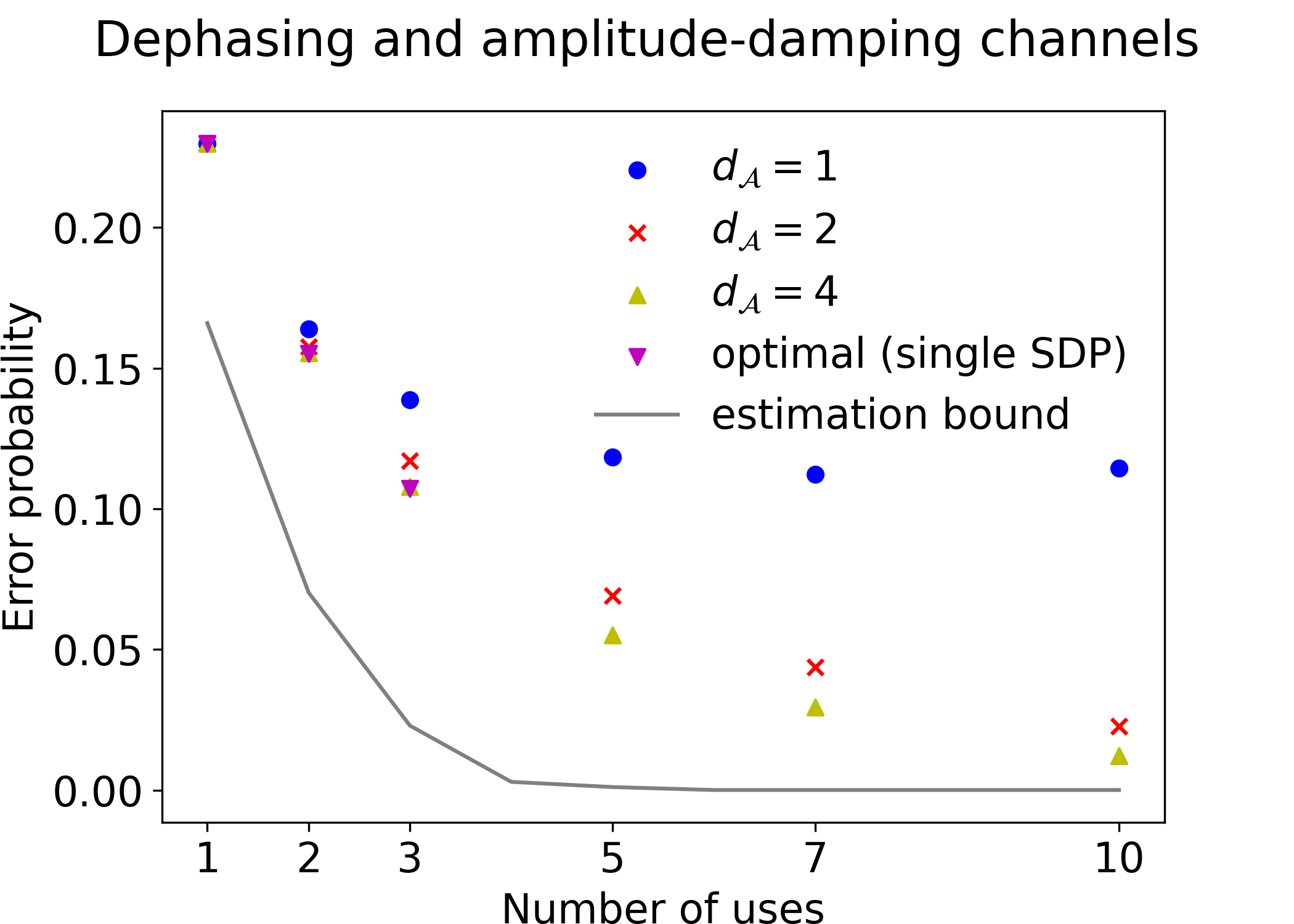}
    \caption{Error probability of discrimination between dephasing and amplitude damping channels for different number of channel uses and ancilla dimensions 1, 2 and 4. Single SDP solution for comparison for 1,2 and 3 uses. The grey solid line shows the estimation bound.}
    \label{fig:bitflip-ampldamp}
\end{figure}

The results of minimal discrimination protocol optimization are presented in Fig.~\ref{fig:bitflip-ampldamp} assuming values of noise parameter to be  $\eta = 0.87$ and $\gamma = 0.67$ (cf. \cite{bavarescoStrictHierarchyParallel2021}). The important observation is a systematic improvement of discrimination procedure with increasing ancilla size. 

Also in Fig.~\ref{fig:bitflip-ampldamp} the estimation bound is presented.
For this model, there is no obvious path between the two channels.
The `straight line' path was chosen -- the QFI was integrated for the parameter $p \in (0,1)$ for a family of channels with such CJ operators:
\begin{equation}\label{eq:convex-sum-channel}
    C(p) = (1-p)  C_1 + p C_2,
\end{equation}
where $C_1,C_2$ are the CJ operators of the two considered channels.
The bound is evidently not tight, even for a few uses, when we are sure of the optimality of the solution.
This means that the bound (at least for the chosen integration path) does not characterise fully the `hardness' of the problem.
The problem of finding the optimal integration path is an interesting one, but goes beyond the scope of the present work.

\section{Conclusions}
\label{sec:conclusions}

The tensor network algorithm presented here works robustly, delivering lower bounds for probability of success for quantum channel discrimination,  even up to 10 and 20 channel uses---a regime inaccessible via full SDP optimization.
It was shown to behave well for the known examples of unitary channels and unitaries with perpendicular noise applied after the signal. The role of ancilla dimension needed to achieve optimal performance has been highlighted.

The results of our algorithm for the considered noise models are close to the presented bounds from estimation.
This not only testifies the effectiveness of the numerical optimization algorithm, but also the usefulness of the bounds derived from quantum estimation theory.
Interestingly, the qualitative character of the initial drop of discrimination error probability with increasing number of channel uses may be related with the properties of the corresponding quantum estimation model (whether it admits the Heisenberg scaling or not), as discussed in \ref{subsec:intuition}.

In the present work, we have considered only the situation in which we are given $N$ copies of a channel.
The presented algorithm can be applied in a straightforward way to discrimination of channels with time-varying parameters: in every use the channel is changed with some time step -- the structure of the algorithm does not necessitate the channels to be equal in every use. 
In principle, this can be made even more general -- to discrimination between any quantum combs that obey some causal relations.
We believe our method could be successfully used in this way for discrimination between classes of time-varying channels.

Finally, let us comment on the important question in channel discrimination is the hierarchy of discrimination strategies (see e.g. \cite{bavarescoStrictHierarchyParallel2021}).
The example due to Harrow et al. \cite{harrowAdaptiveNonadaptiveStrategies2010} is a simple witness of the advantage of adaptive strategies over parallel ones, where all the channels are probed in parallel with a single multi-partite state, which is in the end measured via most general collective measurement. We have tested our the tensor network algorithm with this problem as well, recovering the perfect discrimination adaptive protocol with two channel uses, using single qubit ancilla, as expected. Unlike in quantum metrology, though, it is not clear how to generalize the tensor network approach, to effectively find optimal parallel discrimination schemes, so that one may numerically observe the gap between the best adaptive and best parallel strategies. This has been successfully done in the field of quantum metrology  \cite{chabudaTensorNetworkApproachQuantum2020, dulianQMetroPythonOptimization2025}, using the idea of matrix product states to model input probes and matrix product operators to represent the symmetric logarithmic derivatives needed for computation of quantum fisher information. In the discrimination context, though, we would need to effectively model multi-partite measurements, which are positive operators that need to sum up to identity, the constraints which are challenging to control in the tensor-network optimization framework. We leave this task for a future work.

\emph{Note added.}
Shortly after the completion of the manuscript, a contemporaneous and independent work describing the connection between channel discrimination and estimation appeared in \cite{huangQueryComplexitiesQuantum2025}.
See especially the bound on the discrimination success probability in \cite[Corollary 24]{huangQueryComplexitiesQuantum2025} analogous to our bound in Eqs. \eqref{eq:final-bound-n}, \eqref{eq:qfi_bound}.
We thank the authors for informing us of their related work.

\emph{Acknowledgments.} 
We thank Pavel Sekatski, Francesco Albarelli, Jessica Bavaresco, Marco Túlio Quintino, Łukasz Pawela, Zbyszek Puchała and Mark M. Wilde for fruitful discussion. This work was supported by National Science
Center (Poland) grant No.2020/37/B/ST2/02134

\bibliography{paper-tensors-biblio}



\end{document}